\documentclass[showpacs,aps,pre]{revtex4}
\usepackage{amsmath,epsfig}
\begin{document}
\title{A Hike in the Phases of the 1-in-3 satisfiability}

\author{%
Elitza Maneva$^1$, 
Talya Meltzer$^2$, 
Jack Raymond$^3$, 
Andrea Sportiello$^4$, 
Lenka Zdeborov\'a$^5$}

\affiliation{%
$^1$ University of California at Berkeley, Berkeley, CA 94720, USA
$^2$ School of Computer Science and Engineering, The Hebrew 
University of Jerusalem, 91904 Jerusalem, Israel
$^3$ NCRG; Aston University, Aston Triangle, Birmingham, B4 7EJ
$^4$ Universit\`a degli Studi di Milano, via Celoria 16, I-20133
  Milano
$^5$ CNRS; Univ.~Paris-Sud, UMR 8626, Orsay CEDEX, France 91405, LPTMS
}

\begin{abstract}
We summarise our results for the random $\epsilon$--1-in-3 
satisfiability problem, where $\epsilon$ is a probability of negation 
of the variable. We employ both rigorous and
heuristic methods to describe the SAT/UNSAT and Hard/Easy transitions.  
\end{abstract}


\date{\today}
\maketitle

\section{Introduction}

At the Les Houches ``Complex Systems'' school we learned about spin
glasses and methods by which to study their statistical properties
(G.~Parisi), also the applications of statistical physics to combinatorial
optimization problems were introduced (R.~Monasson). Following the
acquired knowledges we studied the average case of 1-in-3 SAT
problem. We applied rigorous methods to obtain algorithmic and
probabilistic bounds on the SAT/UNSAT and Hard/Easy transitions. We
also employed the cavity method to obtain a more
complete picture of the problem. Our study went beyond a simple
exercise, and led to several interesting results and a separate 
publication of three of the authors~\cite{RSZ07}. 
Here we summarise shortly the methods and give the main results.

1-in-3 SAT is a boolean satisfaction problem. Each formula consists of a set of
variables and clauses, and each clause contains 3 literals. 
A clause is satisfied if exactly one of the literals is True, 
and the formula is \emph{satisfiable} (SAT) if there is any assignment
of variables to True or False, such that every clause is satisfied.
1-in-3 SAT has important similarities to other constraint-satisfaction
and graph-theoretical problems.
It is a canonical NP-complete problem~\cite{cook}
and thus relates to a host of practically relevant problems.
In particular we consider here an ensemble of formulas parameterised
by $\gamma$ and $\epsilon$, where $\gamma$ is the mean connectivity of
variables in the ensemble, and $\epsilon \in [0,1/2]$ is the
probability that variables appear as a negative literal in the
interactions. We then study the SAT/UNSAT and Easy/Hard
transition curves in this space, as the number of variables
$N \rightarrow \infty$. 

This parameterisation is motivated by the knowledge that 
in many related problems there exists a sharp transition in
the typical-case behaviour, from a SAT to an UNSAT regime 
as the parameters are varied;
and a more heuristic Easy-Hard transition, where in a Easy phase
many ``local'' algorithms work in a polynomial time, whereas in the 
Hard phase they need an exponential time.
Previous work in (symmetric) 1-in-3 SAT ($\epsilon=\frac{1}{2}$)
demonstrated that SAT/UNSAT transition is sharp at the threshold
$\gamma^*_{\rm sym} = 1$, and not accompanied by any
Hard region~\cite{ACIM01}.
However, for Exact Cover (i.e.~positive 1-in-3-SAT, $\epsilon=0$) this
threshold is difficult to determine,
with only upper~\cite{KSM04} and lower bounds~\cite{MK05} to the
transition being known, and the presence of a Hard region is suspected.
Studying the threshold behaviour in $\gamma$, for a continuum of
problems parameterised by $\epsilon$, allows us to better understand the
origin of these differences and the nature of the two
transitions.

\section{Results}

\begin{figure}
\begin{center}
\setlength{\unitlength}{30pt}
\begin{picture}(13.5,6)(-0.3,0)
\put(0,0){\includegraphics[scale=1.5]{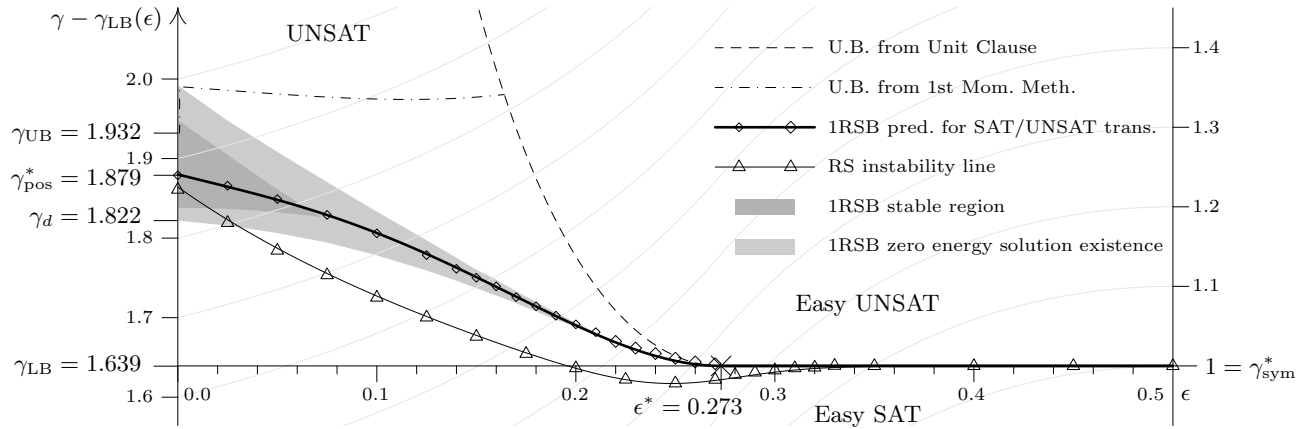}}
\put( 0.58,0.58){\scriptsize 0.0}
\put( 2.83,0.58){\scriptsize 0.1}
\put( 5.34,0.58){\scriptsize 0.2}
\put( 7.84,0.58){\scriptsize 0.3}
\put(10.34,0.58){\scriptsize 0.4}
\put(12.60,0.58){\scriptsize 0.5}
\put(13.15,0.58){$\epsilon$}

\put( 6.25,0.4){$\epsilon^* = 0.273$}

\put(13.45,0.9){$1 = \gamma^*_{\rm sym}$}
\put(13.3,1.95){\scriptsize 1.1}
\put(13.3,2.95){\scriptsize 1.2}
\put(13.3,3.95){\scriptsize 1.3}
\put(13.3,4.95){\scriptsize 1.4}

\put(-0.15,4.55){\scriptsize 2.0}
\put(-0.15,3.55){\scriptsize 1.9}
\put(-0.15,2.55){\scriptsize 1.8}
\put(-0.15,1.55){\scriptsize 1.7}
\put(-0.15,0.55){\scriptsize 1.6}

\put(-1.1,5.3){$\gamma - \gamma_{\rm LB}(\epsilon)$}

\put(0.05,3.29){\makebox[0pt][r]{$\gamma^*_{\rm pos} = 1.879$}}
\put(0.05,2.81){\makebox[0pt][r]{$\gamma_d = 1.822$}}
\put(0.05,3.87){\makebox[0pt][r]{$\gamma_{\rm UB} = 1.932$}}
\put(0.05,0.94){\makebox[0pt][r]{$\gamma_{\rm LB} = 1.639$}}

\put(9.2,0.3){\makebox[0pt][c]{Easy SAT}}
\put(9.2,1.7){\makebox[0pt][c]{Easy UNSAT}}
\put(2.4,5.1){\makebox[0pt][c]{UNSAT}}

\put(8.7,4.95){\scriptsize U.B.~from Unit Clause}
\put(8.7,4.45){\scriptsize U.B.~from 1st Mom.~Meth.}
\put(8.7,3.95){\scriptsize 1RSB pred.~for SAT/UNSAT trans.}
\put(8.7,3.45){\scriptsize RS instability line}
\put(8.7,2.95){\scriptsize 1RSB stable region}
\put(8.7,2.45){\scriptsize 1RSB zero energy solution existence}

\end{picture}
   \caption{\label{phase_diag}%
Phase diagram for the $\epsilon$--1-in-3 SAT problem. On the two axes,
$\epsilon$ and $\gamma$ respectively. In the plot, coordinate
$(\epsilon, \gamma)$ is shifted down by the SCH lower bound 
$\gamma_{\rm LB}(\epsilon)$,
in order to make the lines appear less squeezed. As a consequence, the
Easy-SAT phase w.r.t.~Short Clause Heuristics is the region below the
$x$-axis.}
\end{center}
\end{figure}

Figure~\ref{phase_diag} outlines our present knowledge on the phase
diagram of the $\epsilon$--1-in-3-SAT problem.
Results are consistent with previous statements restricted to the
special cases $\epsilon=0$~\cite{MK05,KSM04}, and 
$\epsilon=\frac{1}{2}$~\cite{ACIM01}.

A rigorous analysis of the \emph{Unit Clause Propagation} (UC) and 
\emph{Short Clause Heuristics} (SCH) algorithms~\cite{ACIM01,MK05} 
led to upper (dashed line) and lower (shifted to $x$-axis) bounds on
the SAT/UNSAT
threshold.  This result identifies two regions in which the problem is
known to be Easy-UNSAT and Easy-SAT, and surprisingly for
$\epsilon>0.273$ the upper and lower bounds coincide, proving the
existence of a range of $\epsilon$ in which, at all values of
$\gamma$, formulas are almost surely Easy.
A second rigorous upper bound for the SAT/UNSAT transition
is obtained by considering the First Moment Method on the 
2-core~\cite{KSM04}, and a third one for the Easy-UNSAT region by an
algorithmic method of embedding formulas in the less
constrained 3-XOR-SAT problem~\cite{fede3xor} 
(out of scale in the figure). Both these bounds are an improvement
w.r.t.~Unit Clause for small values of $\epsilon$, as the UC line
diverges for $\epsilon \to 0$.

The other results on the phase diagram are obtained by the cavity
method \cite{MP99}. We worked both in the assumptions of replica
symmetry (RS: existence of a single pure state) and one-step replica
symmetry breaking~\cite{Parisi80} 
(1RSB: existence of exponentially-many pure states).
Furthermore we checked the stability of the solutions thereby
obtained~\cite{MPR03}.

The RS solution is able to identify the phase transition when $\epsilon$ is
large, but is proven to be unstable at smaller $\epsilon$ (solid line with
triangle marks in fig.~\ref{phase_diag}).  The 1RSB solution we
explored describes clusters which contain solutions (have zero energy)
and disregards their size (entropy)~\cite{MP99}. A 1RSB
solution of this kind 
exists in a region with $\epsilon \lesssim 0.21$, this gives an
indication for the Easy/Hard-SAT transition. The common
interpretation is that the existence of many states (clusters) is
an intrinsic (i.e.~algorithm-independent) reason for the average
computational hardness~\cite{BMW00,MPZ02}. 
The 1RSB solution is shown to be stable in a region with
$\epsilon \lesssim 0.07$, surrounding the SAT/UNSAT curve.
Thus we conjecture that this portion of the curve
is determined exactly.  
In particular, we get for the threshold of the positive 1-in-3 SAT
$\gamma^*_{\rm pos} = 1.879 \pm 0.001$, in accord with existing
exhaustive search results \cite{KSM04,MK05} but with much higher precision.
There exists a region in which neither RS
nor 1RSB assumptions appear to hold sway, 
in this region we can
however guess that 1RSB, as a mean-field assumption, provides an upper
bound to the SAT/UNSAT transition curve~\cite{FL03}.

\section{Concluding Remarks}

Several questions arise out of this study,
also concerning the relation with previous works on 1-in-3 SAT, Exact
Cover and $K$-SAT,  both from Statistical-Physics and algorithmic
perspectives.

The nature of the interactions, being highly constrained (only $3$ out
of $8$ configurations satisfy a clause, and fixing $2$ variables could
violate a clause), leads to remarkable
properties. The first of them is the success of clause-decimation
methods (UC and SCH). The second one, quite unexpected, is the arising
of ``hard contradictions'': above the light gray region in figure,
the solution to 1RSB cavity equations at zero energy becomes singular.

The relative success of the SCH algorithm by comparison with
the RS cavity method is also a novel result.  
We found a region where SCH is able to find a satisfying assignment 
almost surely in polynomial time, and yet from the Statistical-Physics
point of view the replica symmetry is broken. 

Other algorithmic issues should be investigated in the 1-in-3 SAT problem. 
Particularly, the performance of Belief Propagation~\cite{pearl} and Survey
Propagation~\cite{MPZ02} (related resp.~to RS and 1RSB cavity interpretation).
Furthermore, it is interesting to compare our results with the
behaviour of the structurally-affine $(2+p)$-SAT problem~\cite{MZ98}.


\end{document}